\documentstyle[11pt,aaspp4]{article}
\def\ie{{\it i.e.,\ }}
\def\eg{{\it e.g.,\ }}
\def\cf{{\it cf.\ }}
\def\etal{{\it et al.\ }}
\def\msol{\ifmmode {\>M_\odot}\else {$M_\odot$}\fi}
\def\pyr{\ifmmode {\>{\rm\ yr}^{-1}}\else {yr$^{-1}$}\fi}
\def\kms{\ifmmode {\>{\rm km\ s}^{-1}}\else {km s$^{-1}$}\fi}
\def\psqcm{\ifmmode {\>{\rm cm}^{-2}}\else {cm$^{-2}$}\fi}
\def\be{\begin{equation}}
\def\ee{\end{equation}}
\def\bea{\begin{eqnarray}}
\def\eea{\end{eqnarray}}
\received{Feb. 21, 1996}
\accepted{June 12, 1996}
\slugcomment{To appear in The Astrophysical Journal}

\lefthead{Maloney, Begelman \& Pringle}
\righthead{Warping of Accretion Disks}

\begin{document}

\title{Radiation-Driven Warping: The Origin of Warps\\
and Precession in Accretion Disks}

\author{Philip R. Maloney\altaffilmark{1}, Mitchell
C. Begelman\altaffilmark{2,3} and
J. E. Pringle\altaffilmark{4}}

\altaffiltext{1}{Center for Astrophysics and Space Astronomy,
University of Colorado, Boulder, CO 80309-0389; maloney@shapley.colorado.edu}
\altaffiltext{2}{JILA, University of Colorado and National Institute
of Standards and Technology, Boulder, CO 80309-0440; mitch@jila.colorado.edu} 
\altaffiltext{3}{Also at Department of Astrophysical, Planetary and
Atmospheric Sciences, University of Colorado}
\altaffiltext{4}{Institute of Astronomy, Madingley Road, Cambridge CB3
0HA, England}

\begin{abstract}
A geometrically thin, optically thick, warped accretion disk with a
central source of luminosity is subject to non-axisymmetric forces due
to radiation pressure; the resulting torque acts to modify the
warp. In a recent paper, \cite{pri96} used a local analysis to show
that initially planar accretion disks are unstable to warping driven
by radiation torque. Here we extend this work with a global analysis
of the stable and unstable modes. We confirm Pringle's conclusion that
thin centrally-illuminated accretion disks are generically unstable to
warping via this mechanism; we discuss the time-evolution and likely
steady-state of such systems and show specifically that this mechanism
can explain the warping of the disk of water masers in NGC 4258 and
the 164-day precession period of the accretion disk in SS
433. Radiation-driven warping and precession provides a robust
mechanism for producing warped, precessing accretion disks in
active galactic nuclei and X-ray binary systems.

\end{abstract}


\section{Introduction}

An optically thick accretion disk which is illuminated by the central
source of luminosity will reradiate the intercepted flux essentially
normal to the local disk surface. As first noted by \cite{pet77}, if
the accretion disk is warped (\ie has a non-constant inclination
and/or orientation of the line of nodes as a function of radius), then
the resulting radiation pressure force will be non-axisymmetric,
leading to torquing of the disk and modification of the warp. 

The origin of this radiation torque can be seen as follows. Consider
an annulus of 
an accretion disk with a central source of radiation. If we warp this
annulus slightly, it will be illuminated by the central source, and,
if the disk is optically thick, it will absorb the incident flux. The
momentum associated with this absorbed incident flux is purely radial,
and so the force due to the {\it incident} radiation will produce zero
torque. If the disk is optically thick to {\it re-emission} of the
absorbed radiation, however, the net re-emitted flux will be normal to
the local plane of the disk. The radiation pressure acting on the
irradiated surface due to the re-emitted radiation will therefore
exert a torque at each exposed point of the annulus. If the annulus is
warped, \ie the local angle of tilt varies with azimuth, then the net
torque on the ring will be non-zero, because the deposition of
radiation momentum per unit area of the ring will be
non-axisymmetric. In general, the net torque will be non-zero provided
that there is a radial gradient in either the tilt angle $\beta$ (in
which case the ring will precess) or the angle of the line of nodes
(in which case the torque will cause the tilt to change). Note that
the disk must be optically thick both to absorption of the incident
radiation and to re-radiation of the absorbed flux. If the disk is
optically thin to {\it absorption}, then the absorption rate will be
essentially constant through the annulus, and the flux re-radiated
from the two faces of the annulus (the irradiated face and the
opposite face, which does not see the central source directly) will be
the same, so that the re-emitted radiation will exert zero torque. If
the disk is optically thin to {\it re-emission}, then the re-emitted
radiation will be isotropic, and again the torque will be zero.

\cite{pet77} calculated
the influence of this radiation torquing on the disk shape, assuming
an initially warped disk. However, \cite{pri96} has shown that even
initially planar disks are unstable to warping due to radiation
pressure, so that warping should be generic in centrally-illuminated
accretion disks.

\cite{pri96} (hereafter P96) studied the disk dynamics using a local
(WKB-type) analysis, in which the variation of the radiation torque
parameter and viscosity with radius $R$ was ignored. However, such an
approach is only marginally justified, since the unstable modes in
Pringle's analysis have radial wavenumber $k\sim 2\pi R^{-1}$. In this
paper we extend the work of P96 in a global analysis; we show
that this is in fact analytically soluble, with both stable ``normal''
modes and growing modes. In section~\ref{analysis} we solve the
linearized equation for the evolution of the disk tilt, while in
section~\ref{discuss} we compare our results to P96 and
discuss the time evolution and likely steady state of such a
system. We show in section~\ref{appl} that this instability provides a
plausible explanation for the warp observed in the Keplerian disk of
water masers in the nucleus of NGC 4258, and for the 164-day
precession period in SS 433.

\section{Analysis of the Disk Dynamics} \label{analysis}

Following P96, we define $\beta$ to be the
local angle of tilt of the disk with respect to the normal to the
equatorial plane, while $\gamma$ describes the orientation of the line
of nodes with respect to a fixed axis in the equatorial plane.
The equation for the evolution of the local tilt vector, ${\bf
l}(R,t)=(\sin\beta\cos\gamma, \sin\beta\sin\gamma, \cos\beta)$ was
derived by \cite{pri92} (his equation 2.5). We assume an isothermal,
gas pressure-supported disk at $R\gg R_s$ ($R_s$ is the Schwarzschild
radius), for which the surface density $\Sigma\propto R^{-3/2}$ and
the radial velocity $V_R=\nu_1 \Omega^\prime/\Omega$, where primes
denote derivatives with respect to $R$; $\nu_1$ is the
usual disk viscosity, while we denote the viscosity which acts to
reduce the disk tilt by $\nu_2$. This simplifies the evolution equation
considerably (\cf P96), and, assuming that $\beta\ll 1$ and
$\beta'R\ll 1$, it can be written compactly as an equation
for $W\equiv\beta e^{i\gamma}$,
\be
{\partial W\over \partial t}=-i\Gamma {\partial W\over \partial R}
+{\partial\over\partial R}\left({1\over 2}\nu_2 {\partial W\over
\partial R}\right)
\ee
where the radiation torque term is
\be
\Gamma={L\over 12\pi\Sigma R^2\Omega c}
\ee
with $L$ the luminosity of the central source. We make the assumption
that the disk is isothermal, \ie it has the same temperature at all
radii. Then, if the usual Shakura-Sunyaev viscosity parameter $\alpha$
is constant, we find that the viscosity $\nu_1$ scales as
$\nu_1=\nu_o(R/R_o)^{3/2}$ where $R_o$ is an arbitrary fiducial
radius. We further assume that the ratio $\eta=\nu_2/\nu_1$ is
constant. Since the surface density 
$\Sigma=\dot M/3\pi\nu_1$, we can express $\Gamma$ as
\be
\Gamma={\epsilon\over 2\sqrt 2 R_s^{1/2}}{\nu_o\over
R_o^{3/2}}R\equiv\Gamma_o R
\ee
where we have assumed the luminosity is powered by accretion; the
radiative efficiency $\epsilon\equiv L/\dot M c^2$. 
Equation (1) can then be written as
\be
{\partial W\over \partial t}=-i\Gamma_o R {\partial W\over \partial R}
+{1\over 2}{\eta\nu_o\over R_o^{3/2}}{\partial\over\partial
R}\left(R^{3/2}{\partial W\over \partial R}\right)\;.
\ee

\subsection{Particular Solutions}

For illustration we consider two particular solutions of equation (4).

a) The steady-state solution, in which the disk has constant structure
in an inertial frame, can be obtained by setting $\partial W/\partial
t=0$. The general solution for $W(R)$ is then given by
\be
W(R)=c_1 + c_2\int_0^R R^{-3/2}\exp\left({4 i\Gamma_o R_o^{3/2}\over
\eta \nu_o}R^{1/2}\right)\;dR\;,
\ee
where $c_1$ and $c_2$ are arbitrary (complex) constants. Defining 
\be
z={4 i\Gamma_o R_o^{3/2}\over \eta \nu_o}R^{1/2} = i{\sqrt 2
\epsilon\over \eta}(R/R_s)^{1/2},
\ee
(note that $z$ is pure imaginary) this can be written as
\be W(z)=c_1 +c_2'\left [{e^{iz}\over z}+ i\int_z^\infty{e^{is}\over
s}\;ds\right]
\ee
where $c_2'$ is arbitrary. Thus the steady disk has two independent
solutions. The first, $W=$constant, corresponds to the trivial
solution of a flat disk misaligned with the axial plane. The
second solution is (formally) singular at the origin ($W \sim
R^{-1/2}$ as $R\rightarrow 0$). More importantly, we require that the
viscous torque exerted by each annulus of the disk on its neighbor,
given by $G\propto R^{3/2}\partial W/\partial R$, is finite as
$r\rightarrow 0$. The second solution therefore requires a source of
angular momentum at the origin.

b) The `constant-torque' solution is given by
\be
W(R,t)=e^{i\Gamma_o t/2} R^{-1/2}\;.
\ee
This is the solution that results when the viscous torque $G$ is
independent of radius. In this case the disk as a whole precesses with
angular velocity $\Gamma_o/2$. The nodes all lie in the same radial
direction, and the disk is stationary in a frame which rotates with
angular velocity $\Gamma_o/2$. This comes about because the radial
dependence of $W$ insures that the precession rate induced at each
radius is independent of radius (\cf equation [3.10] in P96). 

This is actually a special case of a more general class of solutions,
$W(R,t)=e^{i\Gamma_o t/a} R^{-1/a}$,
where $a \neq 0$. The difference is that for these solutions
$\eta=\nu_2/\nu_1$ is not constant, as $\nu_2\propto R^{1+1/a}$. 

\subsection{More General Solutions}

In order to investigate more general time dependence, we Fourier
transform with respect to time and write $\partial W/\partial t=
i\sigma W$; $\sigma$ in general has both real and imaginary parts. We
again make the substitution (6) to rewrite equation (4) in the form
\be 
z{\partial^2 W\over \partial z^2} + (2-z){\partial W\over\partial
z} -{2\sigma\over \Gamma_o}W=0.
\ee 

This is Kummer's equation (\eg \cite{abs64}, (AS), equation 13.1.1),
$zW''+(b-z)W'-aW=0$, with $b=2$ and $a=2\sigma/\Gamma_o$. The solutions
to Kummer's equation are confluent hypergeometric functions,
$M(a,b,z)$ and $U(a,b,z)$. The condition that zero torque is applied
to the disk at the origin now becomes $z^2 W'\rightarrow 0$ as
$z\rightarrow 0$. However, as $|z|\rightarrow 0$, $U\sim |z|^{1-b}$
and thus the contribution of the Kummer $U$ function provides a finite
torque at the origin. For physical reasons we exclude the $U$ function
from the solution.  Thus the spatial part of $W$ is just given by the
Kummer $M$ functions. Using the asymptotic expression for $M(a,b,z)$
(AS 13.5.1), we find that imposing the condition $W(z)\rightarrow 0$
as $z\rightarrow \infty$ requires that the real part of $a$ satisfy
\be 0 < {\rm Re}(a) < 2\;,\qquad {\rm that\ is,}\qquad 0 < {\rm
Re}(\sigma) < \Gamma_o\;.\ee 
This constraint on Re$(a)$ is also required if the
intercepted flux $dL$ (see equation (2.13) of \cite{pri96}) is to
remain finite as $z\rightarrow\infty$.  It is not possible to impose
the stronger regularity condition that $|z^2W|\propto |RW|=|\beta R|$
(\ie the height of the warp above the untilted equatorial plane)
remain finite as $z\rightarrow\infty$, as $z^2W$ diverges for all
Re($a$). Real astrophysical disks will always have finite $RW$,
however, either due to physical truncation of the disk or because the
surface density declines to the point where the disk becomes optically
thin to the bulk of the incident flux.
 
A particular case of interest occurs when Re$(a)=1$, that is,
$a=1-i\omega$, with $\omega$ real. This corresponds to a steady
precession of the disk structure with angular velocity $\Gamma_0/2$,
and an exponential growth of disk structure at a rate $e^{\omega
t}$. In this case the $M$ functions are simply
related to the $L=0$ Coulomb functions:
\be
M(1-i\omega,2,2ix)=e^{ix} F_0(\omega,x)x^{-1}/C_0(\omega)
\ee
(AS 13.6.8) where $\omega$ can be either positive or negative and
$F_0$ is the regular Coulomb wave function with $L=0$ (AS 14.1.4; note
that both $\omega$ and $x$ are real). The azimuth of the line of nodes
is given by $\gamma=x$, \ie $\gamma\propto R^{1/2}$, which corresponds
to a prograde spiral (\cf P96). 

If $\omega=0$, there is no growth and the disk structure is steady in
the precessing frame. In this case $M$ simplifies to the solution
\be
M(1,2,2ix)=e^{ix}{\sin(x)\over x}\;.
\ee
It is interesting to compare this solution (zero torque at the origin)
with the torqued solution, equation (8), found above. Both solutions
precess at the same rate, $\Gamma_o/2$, and both solutions display an
envelope to $|W|$ of the functional form $R^{-1/2}$. In both cases,
the precession is driven solely by radiation (and thus the precession
rate depends only on $\Gamma_o$). However, the zero torque solution
maintains its constant shape (in the precessing frame) by using
viscous forces, which would tend to smooth out the oscillations of
$\beta$ with radius, to balance the radiation induced growth which
comes about because $\partial \gamma/\partial R > 0$ (\cf P96,
equation [3.10]).

Some general points regarding the $M$ functions are worth
mentioning. As $z\rightarrow 0$ $M(a,b,z)\rightarrow 1$ for all $a$,
$b$, indicating that the disk has a finite tilt at the origin. Note
that since we have required Re$(\sigma) > 0$, the tilt is not
stationary in the inertial frame ($\sigma=0$ is discussed in section
2.1). As we will see below, however, the tilt at $R=0$ for the growing
modes can be extremely small. The Re$(a)=1$ $M$ functions are the only
genuinely oscillatory (\ie zero-crossing) functions; the radial part
$|M|$ is symmetric about $a=1$ for $0 < {\rm Re}(a) < 2$ but the
angular dependence differs for Re$(a)> 1$ and Re$(a) < 1$. Figure 1
shows $|M|\propto |\beta|$ as a function of $|z|$ for several values
of Re($a$) with ${\rm Im}(a)\equiv-\omega=0$. For $\omega> 0$
(corresponding to growing modes), the first zero (or minimum) of $|M|$
moves to larger $|z|$, and there is a dramatic increase in the
amplitude of the first maximum of $|M|$, which also shifts to larger
$|z|$. Both the magnitude of the shift and the increase in amplitude
increase with $\omega$. For example, Figure 2 plots $|M|$ for
Re$(a)=1$ and several values of $\omega$; the behavior for other
values of Re$(a)$ is very similar. All the $|M|$ have been normalized
to a maximum of one, and are labeled with $\omega$ and the
normalization constant $\beta_o$. For $\omega=10$, the normalization
constant $\beta_o\approx 2.5\times 10^{-12}$.

As we show below, the maximum value $\omega_{\rm max}$ of $\omega$
(and therefore of the 
growth rate $\sigma$) is constrained by the outer radius of the disk
(either the physical edge or the optically thick edge). There are,
however, likely to be unstable modes present throughout the range $0 <
\omega < \omega_{\rm max}$. This can be seen by considering a WKB-type
analysis and replacing $\partial/\partial R$ by $ik$ (\cf P96). In
this case we obtain 
\be
\sigma = i \left\lbrace -\Gamma_o kR + {{1\over 2}\nu_o (kR)^2\over
R^{1/2}} \right\rbrace\;.
\ee
Thus for a fixed $kR$ (\ie a fixed radial mode structure) we see that
$\omega\propto -{\rm Im}(\sigma)$ is largest, $\omega_{\rm max}$, at
large radius. Conversely, as $R$ decreases it is evident that all
unstable values of $\omega$ in the range $0 < \omega < \omega_{\rm
max}$ are present. Physically, this arises because
the viscosity ($\nu\propto R^{3/2}$) has a stronger radial dependence 
than the radiation torque parameter ($\Gamma\propto R$). Thus
for arbitrarily small $\omega$ there is always some radius at which
the torque term dominates over the viscous term, and the disk is
unstable to warping. In a real system, however, the development of
small-scale modes will be limited by the disk age, as we discuss 
below.
 
\section{Time Evolution and Steady State} \label{discuss}

\cite{pri96} concluded that centrally-illuminated accretion disks are
unstable to warping beyond a critical radius $R_c$, given by
\be
{R_c\over R_s}=\left(2\sqrt 2\pi\eta\over\epsilon\right)^2\;;
\ee
the disks remain flat interior to this radius. At first glance this
appears substantially different from the results of
section~\ref{analysis}, which showed that the tilt $\beta$ must be
non-zero at the origin. However, the time evolution of the tilt is
crucial. With increasing $\omega$, the initial growth of the
warp occurs at larger radius, and the amplitude of the maximum of
$\beta$ relative to the $R=0$ value grows dramatically (\cf Figure 2);
for Re$(a)=1$, $\beta_{\rm max}/\beta_o\propto
(\exp(2\pi\omega)-1)/\omega$. Thus the disk will begin to warp
initially at the largest possible radius (close to the disk physical
or optical depth edge) and will be flat interior to this radius, as in
Pringle's analysis. The main difference is that here the warp has
non-zero precession frequency (\ie the real part of $a$ is non-zero),
whereas in P96 the growing modes were purely imaginary. Note
also, from equation (6) for $z$, that if the warp begins at some value
of $|z|=|z|_w$, this corresponds to a radius
\be
{R_w\over R_s}={1\over 2}\left(\eta\over\epsilon\right)^2 |z|_w^2
\ee
which shows the same scaling with $\eta$ and $\epsilon$ as equation
(4.3) of P96, given above.

If we require that the disk be unwarped beyond some maximum radius
$R_{\rm max}$ (as will be the case, for example, if the disk goes
optically thin beyond some radius, so that the instability ceases to
operate), then we must have ${\rm Re}(a)=1$, as this is the only mode
that passes through zero. The warp then has a unique precession rate,
${\rm Re}(\sigma)=\Gamma_o/2$. Furthermore, the requirement that the
first zero of $\beta$ (and hence $|M|$) occur at $R_{\rm max}$ fixes
a unique value for the the maximum value of $\omega$ (\cf Figure
2). (A similar boundary condition will determine $\omega_{\rm max}$
even if $a\neq 1$, since the radius of the warp onset always increases
with $\omega$.)

One might expect that the steady state of such an accretion disk will
be the $\omega=0$ solution. However, real accretion disks will not
necessarily reach this state. The tilt $\beta$ -- and
therefore the amplitude of the warp -- will saturate at some finite
value, due to nonlinear effects which have been ignored in our
analysis. This will of course occur first for the fastest growing
mode, at the largest unstable radius. With time the slower-growing
modes (only those with zeros at $R_{\rm max}$ if we require the disk
to be flat there) will develop at smaller radius. However, due to the
rapid reduction in amplitude $\beta_{\rm max}/\beta_o$ with decreasing
$\omega$, the slower-growing modes have to evolve through a
progressively larger number of $e-$folding times to reach the
amplitude of the mode with $\omega=\omega_{\rm max}$. For Re$(a)=1$,
the number of required $e-$folding times is approximately
$\pi(\omega_{\rm max}-\omega)$ (assuming $2\pi\omega\gg 1$). Thus the
minimum value of $\omega$ that can have developed significant
amplitude is set approximately by the condition 
\be
\tau_{\rm disk}={\pi(\omega_{\rm max}-\omega_{\rm
min})\over\omega_{\rm min}}\tau_i(\omega_{\rm max})
\ee
where $\tau_{\rm disk}$ is the lifetime of the accretion disk and
$\tau_i(\omega_{\rm max})$ is the timescale for growth of the mode
with $\omega=\omega_{\rm max}$. As we
will see in section~\ref{appl}, this is likely to be pose a
significant constraint on $\omega_{\rm min}$ for accretion disks
around supermassive black holes in active galactic nuclei. 

In addition, the effects of shadowing -- which we have ignored -- may
be important for modes which have multiple zeros between $R=0$ and
$R_{\rm max}$. This will depend on both the actual size of the central
source of luminosity and the amplitude at which the warp eventually
saturates.

Figure 3 shows the shape of the warp for the $a=1$, $\omega=0$
(steady-state) solution. The disk surface has been plotted out to the
radius ($x=2\pi$) at which the warp returns to the equatorial plane. The
maximum amplitude of the warp has been fixed at 10\% of this
radius. The spiral shape of the line of nodes is readily visible.

\section{Applications} \label{appl}

\subsection{The Warped Maser Disk of NGC 4258}

VLBI observations of the 22 GHz water maser line emission from the
nucleus of NGC 4258 (which possesses a low-luminosity active nucleus)
have shown that the maser emission arises in a thin, warped disk, with
an inner radius of 0.13 pc and an outer radius of 0.25 pc,
which exhibits a perfectly Keplerian velocity curve to within the
measurement errors (\cite{gre95}; \cite{miy95}). The derived central
mass is $M_c=3.6\times 10^7\msol$, and as shown by \cite{mao95},
dynamical arguments essentially rule out mass models other than a
supermassive black hole. The obliquity of the warp is $\mu\approx 0.25$
at the outer edge of the masing zone (Greenhill 1995, private
communication) and decreases steadily inward.

\cite{nmc95} showed that X-ray irradiation of dense molecular gas is
an effective mechanism for generating powerful water maser emission;
this also naturally explains the association of water megamasers with
active galactic nuclei (\cite {cla86}; \cite{bra94}). These models
also show that there is a critical pressure at which the gas goes from
molecular to atomic. By modeling the NGC 4258 disk using the standard
$\alpha-$prescription for viscosity, \cite{neu95} (NM95) showed that
the disk would be extremely thin ($h/R\ll 10^{-2}$), that the
transition from molecular to atomic gas should be identified with the
outer edge of the masing region, and that the mass accretion rate
through the disk is $\dot M/\alpha\approx 7\times 10^{-5}\msol\pyr$.
This mass accretion rate in turn implies that the radiative efficiency
$\epsilon\approx 0.1$. The existence of the warp in this picture is
crucial, since for a flat disk neither the greatly reduced irradiation
from the central source nor heating by viscous dissipation would keep
the disk warm enough to produce substantial amounts of water or excite
the masing transitions. In fact, NM95 suggested that the inner
edge of the masing disk marked the inner boundary of the warp, as is
the case in the model for the warp derived by \cite{miy95}. Although
an arbitrary imposed warp would be kinematically stable in a Keplerian
potential, since the vertical and azimuthal frequencies are the same, the
origin of the warp in NGC 4258 remained puzzling.

Here we show that the radiation-driven warping instability discovered
by \cite{pri96} provides a plausible explanation for the warping of
the disk. From equation (6) for $z$, with $R_s\simeq 1.1\times
10^{13}$ cm for the central object in NGC 4258, we find that the inner
and outer edges of the masing region are at
\be
|z|_{\rm in}\approx 28\epsilon_{0.1}\;,\qquad |z|_{\rm out}\approx
39\epsilon_{0.1} 
\ee
where $\epsilon=0.1\epsilon_{0.1}$. As can be seen from Figure 2, this
requires $\omega_{\rm max}\sim 10$ for $\epsilon_{0.1}\sim 1$, as is
indicated by the results of NM95. (We note that in the molecular zone
at the outer edge of the masing region in the model of NM95, the
hydrogen column density through the disk $N_H\sim 10^{24}\psqcm$,
satisfying the assumption that the disk is optically thick to both
absorption and emission. Also, the outer edge of the masing zone does
not have to coincide with the maximum of the warp.) It is likely that
$\omega_{\rm min}$ is not much smaller than this value, due to the
limit set by the finite age of the accretion disk. The characteristic
timescale for growth of the instability is \be
\tau_i(\omega)\approx{1\over |\sigma|}={2\over \omega\Gamma_o} \ee
assuming that $\omega\gg {\rm Re}(a)$. This can be expressed in terms
of the viscous timescale $\tau_{\rm visc}\sim 2\pi R^2\Sigma/\dot M$:
\be \tau_i(\omega)\sim {12\over \omega}|z|^{-1}\eta^{-1}\tau_{\rm
visc} \ee which, under the assumption that the luminosity is
accretion-powered, depends only on $\epsilon$ and not on $L$ and $\dot
M$ individually. Assuming standard $\alpha-$viscosity and negligible
self-gravity for the accretion disk, this becomes \be
\tau_i(\omega)\sim{4\over \omega}{c R_s\over \alpha c_s^2\epsilon}
\approx 3\times 10^7 {M_8\over \omega \alpha T_3 \epsilon_{0.1}}\;{\rm
yrs} \ee where $M_c=10^8 M_8\msol$, $c_s$ is the sound speed, the gas
temperature $T=10^3 T_3$ K, and the numerical coefficient is for
molecular gas. In the masing molecular zone, $0.3 < T_3 < 1$. While
the characteristic growth time for the $\omega=10$ mode in NGC 4258 is
$\tau_i\sim 5\times 10^6$ years, the time required for even the
$\omega=5$ mode to become comparable in amplitude is an order of
magnitude longer. Thus we expect that the warp will be dominated by
modes with $\omega$ close to $\omega_{\rm max}$, unless the accretion
disk is extremely long-lived ($\tau_{\rm disk}\gg 10^8$ years). Hence
the predictions of the radiation-driven warp instability model are in
reasonable agreement with the observations of NGC 4258, for
$\epsilon\sim 0.1$.

This instability cannot provide an explanation for the warp
if NGC 4258 possesses an advection-dominated accretion disk with
$\epsilon \sim 10^{-3}$, as suggested by \cite{las96}.
From equation (17), this requires that $|z|_{\rm in}$,
$|z|_{\rm out}\ll 1$. However, the only modes that have any power at
such small $|z|$ have $\omega < 1$. Equation (20) shows that the
timescale for growth of these modes is prohibitively long, with
$\tau_i > 10^9$ years. 

\subsection{The 164-day Precession Period of SS 433}

SS 433 is an eclipsing binary system containing an early-type
star and an accretion disk around a compact object. The optical and
infrared spectrum shows two sets of highly Doppler-shifted emission
lines, which reach velocity extremes of $+50,000$ and $-30,000$ \kms,
respectively. These emission lines are believed to arise from
oppositely-directed relativistic ($v/c\approx 0.26$) beams of
material emanating from SS 433 (reviews of observations and models of
SS 433 can be found in \cite{mar84} and \cite{zwi89}). The
high-velocity systems show systematic changes in wavelength, with a
period of approximately 164 days. This period is generally regarded to
be the result of precession of the ejection axis of the system. There
is also direct evidence for precession of the beams from radio maps
(\cite{hje86}). In addition, there is evidence for precession of the
accretion disk, which dominates the optical light of the system, from
optical photometry (\cite{mar84}). 

A number of models have been proposed to explain the precession of the
accretion disk. Most of these involve ``slaved'' precession, in which
the residence time in the disk is sufficiently short that the inner
disk follows the precessional motion of a misaligned companion star or
outer disk (\eg \cite{kat82}). This in turn requires that the disk be
very thick, if the internal disk pressure is to be comparable to the
very large viscous stresses which are required. Support for the
existence of a thick ($h/R\sim 1$) disk is provided by analysis of the
light curve (\cite{and83}).

However, if the accretion disk in SS 433 is substantially warped, the
actual thickness of the disk may be much smaller than inferred from
the light curve, which will measure the amplitude of the warp. (This
is certainly the case in NGC 4258, in which the solid angle occupied
by the disk about the source is completely dominated by the warp.) As
shown in section~\ref{analysis}, the real part of $a$ is in general
not zero, so the warp will precess with a fixed pattern
speed. Assuming that the accretion disk in SS 433 is subject to the
radiation-driven warp instability, we now show that that the 164-day
precession period can be explained as the result of precession of the
warp. 

We again assume that Re$(a)=1$, so that we can require the outer
boundary of the disk to be unwarped. As noted earlier, this fixes the
precession rate at $\Gamma_o/2$ and so the precession timescale is
just $T_{\rm prec}\approx4\pi/\Gamma_o$. Since the assumption of an
isothermal disk is likely to be poor for SS 433, we instead write the
viscosity in the form
\be
\nu_1={\alpha\over \sqrt 2} cR \left({R\over R_s}\right)^{-1/2} \left
({h\over R}\right)^2
\ee
and so the precession period is
\be
T_{\rm prec}\approx {16\pi\over \alpha\epsilon}{R\over c}\left({R\over
h}\right)^2
\ee
where this should be evaluated at $R=R_{\rm warp}$. For accretion
disks around $M_c\sim$ solar-mass objects, the timescales for growth of
the unstable modes are sufficiently short that the warp reaches the
steady-state, $\omega=0$ solution. For $a=1$ the first maximum occurs
at $|z|\approx 1.3\pi$, and we take this to correspond to $R_{\rm
warp}$. Thus the precession timescale becomes
\be
T_{\rm prec}\sim 430{\eta^2\over \alpha\epsilon^3}{R_s\over c} \left(
{R\over h}\right)^2_{\rm R_{\rm warp}}\;.
\ee
For SS 433, $\epsilon$ is probably $\ll 0.1$, since the accretion rate
is almost certainly substantially super--Eddington (\eg \cite{beg80})
and so the precession timescale is
\be
T_{\rm prec}\sim 50{\eta^2\over\alpha\epsilon^3_{0.001}}
\left({M_c\over\msol}\right)\ \left({R\over h}\right)^2_{\rm R_{\rm
warp}}\;{\rm days}
\ee
which can explain the 164-day precession period in SS 433 for
reasonable values of $\alpha$ and $(R/h)$. We also note that the
precession can be either prograde or retrograde with respect to the
direction of disk rotation, as can be seen from the derivation leading
to equation (2.18) of P96.

\section{Summary}

Our global analysis confirms the conclusion of \cite{pri96} that thin,
centrally-irradiated accretion disks are generically unstable to
warping driven by radiation pressure. The maximum growth rate is set
by the largest unstable radius (either the physical boundary of the
disk or the radius at which it becomes optically thin to the bulk of
the incident or re-emitted flux). The growth rate on smaller scales in
accretion disks around supermassive black holes in active galactic
nuclei will in general be limited by the lifetime of the accretion
disk. Although the final state of the instability will be determined
by nonlinear processes which have not been considered in this
analysis, we have shown that the warping instability provides an
explanation for the existence of the warp in the masing accretion disk
in NGC 4258. Furthermore, for plausible parameters the 164-day
precession period in SS 433 can be identified with the precession
period of the warp, as can the long ($30-300$ day) periods seen in
other X-ray binary systems, such as Her X-1 (see \cite{whi95} for a
review).

Radiation-driven warping and precession thus offers a robust mechanism
for producing warped, precessing accretion disks in AGN and in
accreting binary systems such as Her X-1 and SS 433. Because the warp
adjusts itself in such a way as to precess at a uniform rate at all
radii, radiation-driven warping is an inherently global mechanism,
thereby avoiding the difficulties inherent in other proposed
mechanisms for producing warping and precession. This instability
should generally be important in thin accretion disks.

\acknowledgments 

PRM acknowledges support through the NASA Long Term Astrophysics
Program under grant NAGW-4454. MCB was supported by NSF grant AST
91-20599.

\clearpage

%\begin{figure}
\figcaption[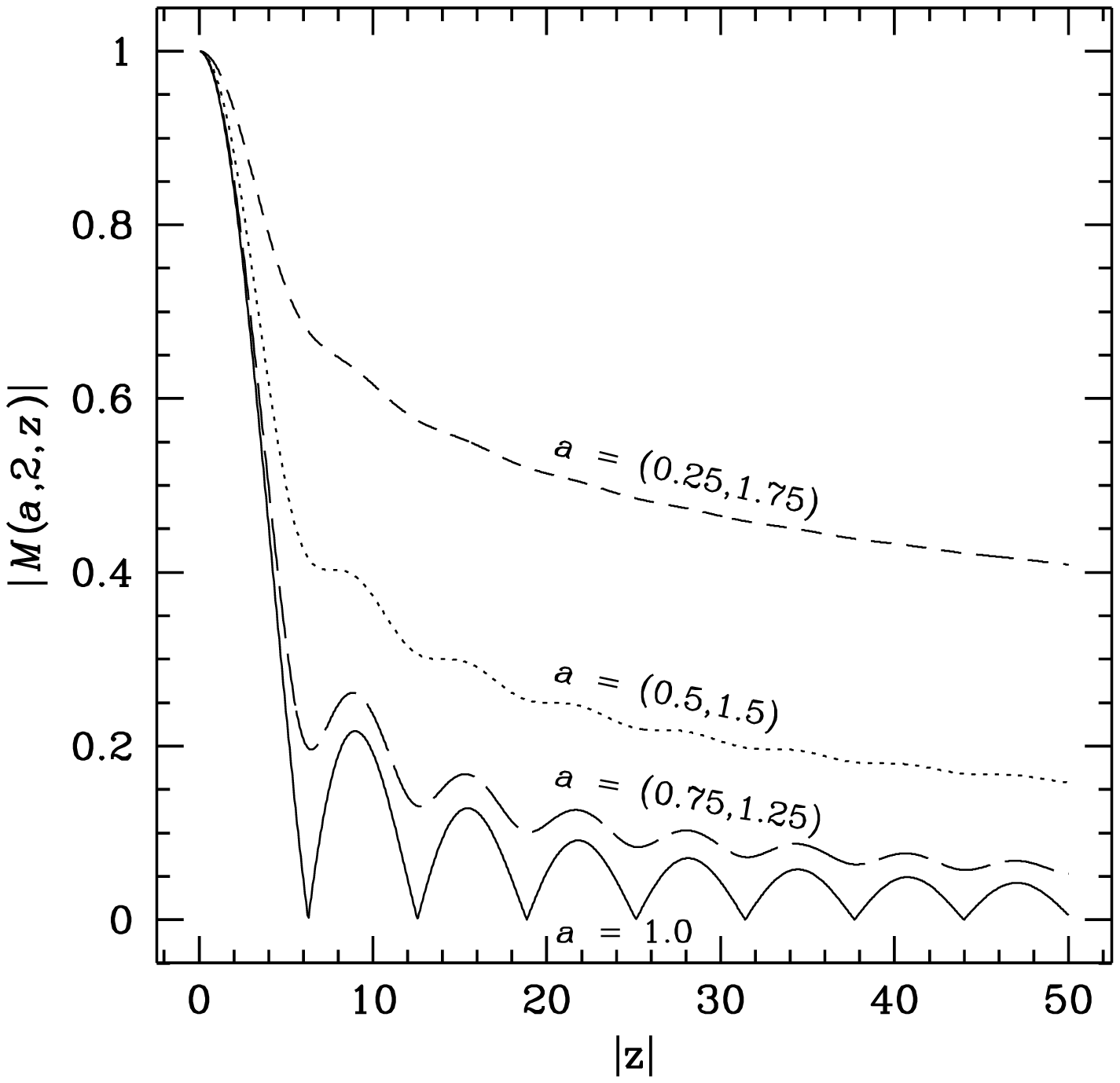]{The absolute magnitude of the Kummer function
$M(a,2,z)$ 
for $\omega=-{\rm Im}(a)=0$ and several values of Re($a$) (\ie the 
time-independent solutions of the twist equation). The
magnitude of the local angle of tilt $|\beta|$ is equal to $|M|$ times
a normalization coefficient. The radial variable $z$ is defined by
$z=i\protect\sqrt 2\epsilon (R/R_s)^{1/2}/\eta$, where $\epsilon$ is the 
radiative efficiency $\epsilon=L/\dot M c^2$, $R_s$ is the
Schwarzschild radius, and $\eta$ is the ratio of vertical to radial
viscosities. The Kummer function with Re$(a)=1$ is the only
zero-crossing function. For other values of $a$ $|M|$ is symmetric
about $a=1$. Requiring that the tilt $\beta\rightarrow 0$ as
$R\rightarrow\infty$ requires that the real part of $a$ satisfy $0 <
{\rm Re}(a) < 2$.}
%\end{figure}

%\begin{figure}
\figcaption[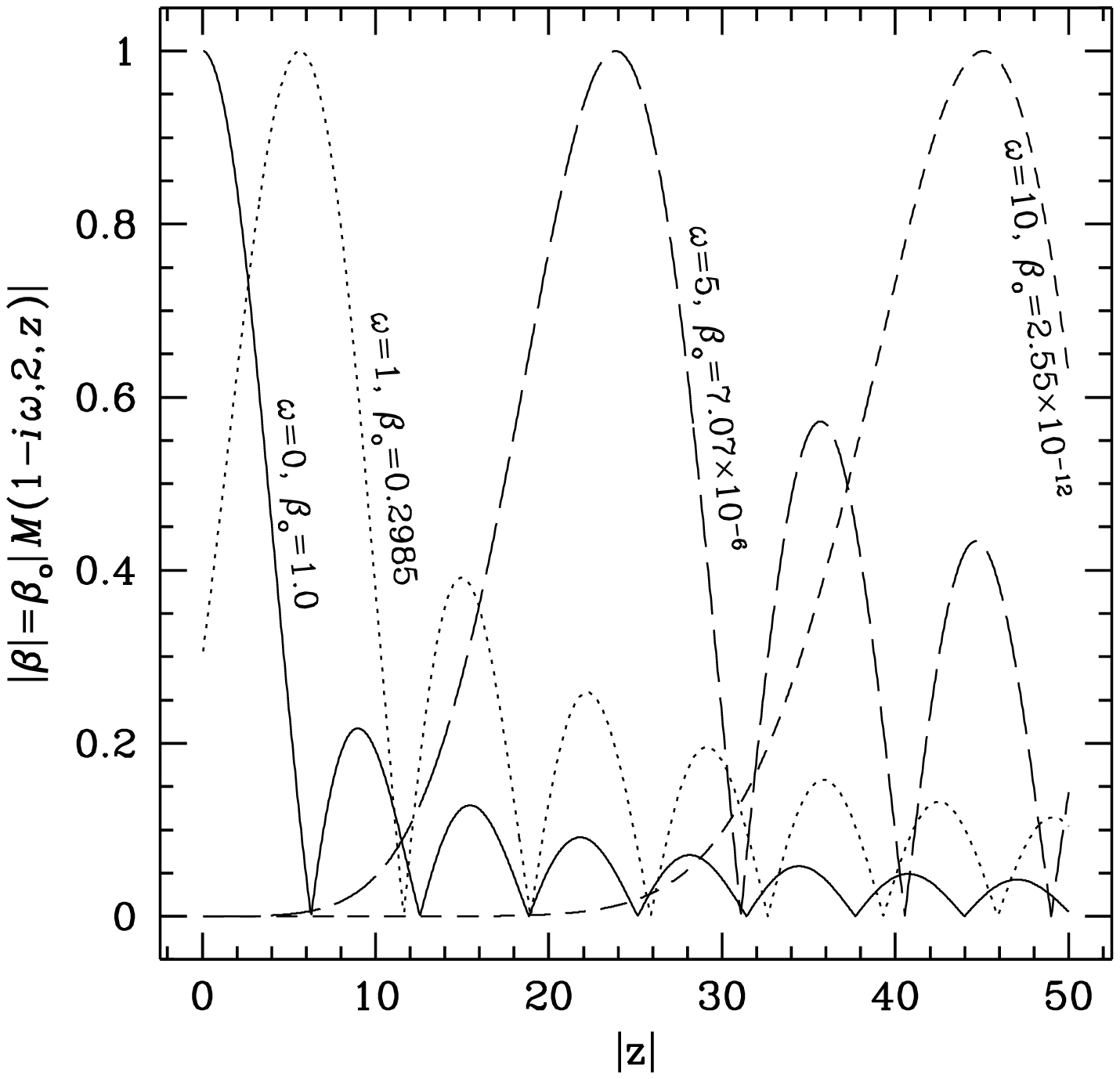]{The amplitude of the local angle of tilt $|\beta|$ for
several of the growing solutions of the twist equation. The Kummer
functions $|M(1-i\omega,2,z)|$ are plotted. All these modes have
Re$(a)=1$ and grow as $e^{\omega t}$. They have all been normalized so
that the maximum amplitude is 1; each curve is labeled with $\omega$
and with the normalization constant $\beta_o$, \ie the magnitude of
the tilt at the origin. For $\omega\gg 1$ this is extremely small, so
that for modes with high growth rates the disks remain flat interior
to a critical radius. The maximum value of $\omega$ for a disk is
fixed by the largest unstable radius, which is either the physical
edge of the disk or the radius at which it becomes optically thin.}
% (see
%section~\ref{discuss}).}
%\end{figure}

\figcaption[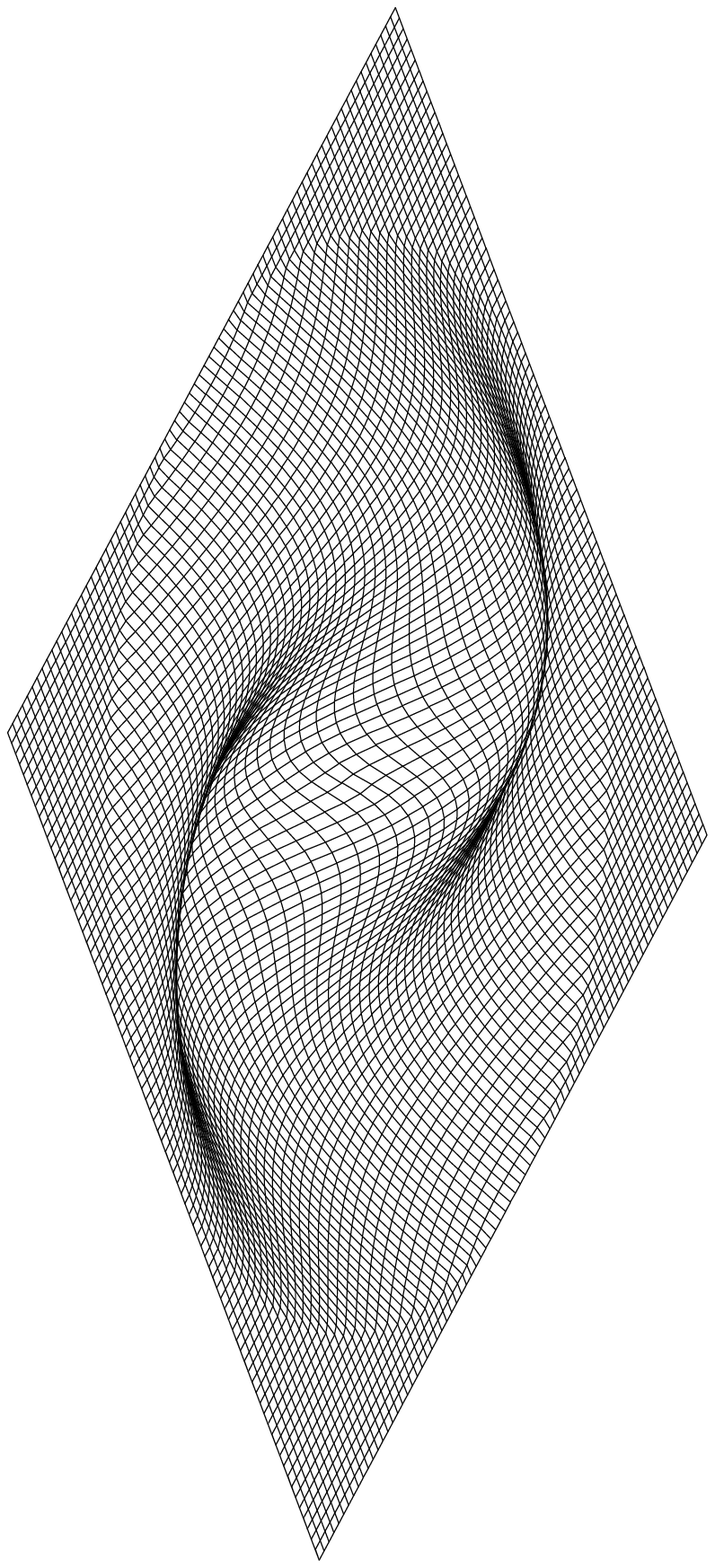]{Surface plot of the warp shape for the
steady-state ($\omega=0$), $a=1$ mode of an isothermal accretion
disk. The warp has been plotted out to the radius at which it returns
to the equatorial plane; the amplitude of the warp has been fixed at
10\% of this radius. }
\clearpage

\plotone{figure1.eps}

\clearpage

\plotone{figure2.eps}

\clearpage

\plotfiddle{figure3.eps}{8in}{0}{100}{100}{-288pt}{-144pt}

\end{document}